\documentclass{jfm}

\pdfoutput=1

\input{jfmdefs}

\usepackage{amsmath}
\usepackage{bm}
\usepackage{subfigure}
\usepackage{graphicx}
\usepackage{booktabs} 
\usepackage{natbib}
\usepackage{afterpage}
\usepackage{tikz}
\usepackage{xcolor}
\newcommand{\url}[1]{\texttt{#1}}
\graphicspath{{figs/}}

\newcommand{\mathnotation}[2]{\newcommand{#1}{\ensuremath{#2}}}

\renewcommand{\time}{t}
\mathnotation{\ldef}{\mathrel{\raisebox{.069ex}{:}\!\!=}}
\mathnotation{\rdef}{\mathrel{=\!\!\raisebox{.069ex}{:}}}
\mathnotation{\ee}{\mathrm{e}}              
\mathnotation{\dint}{\,{\mathrm{d}}}        
\mathnotation{\xv}{\bm{x}}                  
\mathnotation{\urh}{u_\rho}                 
\mathnotation{\uz}{u_z}                     
\mathnotation{\ruv}{\hat{\bm{r}}}           
\mathnotation{\ptl}{\partial}               
\mathnotation{\Uc}{U}                       
\mathnotation{\pal}{\lambda}                
\mathnotation{\Dist}{L}                     
\mathnotation{\lsc}{\ell}                   
\mathnotation{\Lsep}{L}                   
\mathnotation{\nd}{n}                       
\mathnotation{\ai}{a}                       
\mathnotation{\bi}{b}                       
\mathnotation{\dadb}{\dint\ai\dint\bi}      
\mathnotation{\da}{\dint\ai}                
\mathnotation{\db}{\dint\bi}                
\mathnotation{\Vol}{V}                      
\mathnotation{\Nenc}{M}                     
\mathnotation{\Deltarvar}{\delta}           
\mathnotation{\C}{C}                        
\mathnotation{\Rdisk}{R}                    
\mathnotation{\Displacement}
{\Delta_\lambda(a,b)}                     
\mathnotation{\DisplacementSquared}
{\Delta_\lambda^2(a,b)}                   
\mathnotation{\chasing}{\theta=0}          
\mathnotation{\tilting}
{\theta=\pi/4}                    
\mathnotation{\sweeping}
{\theta=\pi/2}                    
\mathnotation{\kappas}{\kappa_{\text{s}}}     

\mathnotation{\derd}{\text{d}}      
\mathnotation{\Rep}{\text{Re}}    
\mathnotation{\Imp}{\text{Im}}    

\title[Stirring by multiple cylinders in potential flow]{Stirring by multiple cylinders in \\potential flow}

\author[Z. Lin and Y. Zhang]%
{
  Z\ls H\ls I\ns
  L\ls I\ls N$^1$
  \and
  Y\ls U\ls A\ls N\ls Z\ls H\ls A\ls O\ns
  Z\ls H\ls A\ls N\ls G$^2$
}

\affiliation{
  $^1$ School of Mathematical Sciences, Zhejiang University, Hangzhou, Zhejiang 310027, China\\
  $^2$ Department of Engineering Sciences and Applied Mathematics, Northwestern University, Evanston, Illinois 60208-3125, USA\\
}

\begin{document}

\maketitle

\begin{abstract}
We consider the enhanced mixing due to multiple cylinders organised in schools moving synchronously in a potential flow.  Here simple interactions between cylinders are modelled by the method of image doublets.   This is an extension to Thiffeault \& Childress's work [\emph{Physics Letters A} \textbf{374}, 3487 (2010)] where fluid particle displacements due to non-interacting swimmers were analysed to produce an effective diffusivity that may have a significant impact in ocean mixing.    Our results show that schools of two cylinders induce nonlinearly boosted diffusivity compared with the non-interacting case for general configuration parameters, except when they move along a straight line with small separation.   We attribute this phenomenon to two different physical mechanisms via which interacting cylinders cooperate to generate long particle drifts depending on their formation.  Finally, the effective diffusivity of schools of three or more cylinders in various configurations are also discussed. 
\end{abstract}

\section{Introduction\label{sec:intro}} 

Extensive study has been evoked in the last decade about the biogenic impact on ocean mixing due to the swimming motions of marine organisms.  In efforts to support or disprove the significance of such an input first proposed by \cite{Munk1966}, several studies offered partial yet inconclusive arguments from the perspective of energy budget and efficiency \citep{Dewar2006, Huntley2004, Kunze2006, Leshansky2010, Underhill2008, Visser2007, Wagner2014}.  As the scientific debate continues, the complex nature of the problem requires better understanding in the behaviours and characteristics of marine swimmers, and in the mechanisms that couple small-scale swimming and large-scale mixing \citep{Katija2011}.    

\cite{Katija2009} suggested that Darwinian drift \citep{Darwin1953} is one such mechanism that could result in enhanced mixing.  Thiffeault and Childress (2010) proposed a stochastic hydrodynamics model, in which the swimming bodies form a dilute suspension of cylinders or spheres that move in random directions.    Consequently, an integral formula for the effective diffusivity in a potential flow or in a Stokes flow with slip boundary conditions was derived and was verified by numerical simulations \citep{Thiffeault2010b, Lin2011}.  With physical parameters, the theoretical prediction implies a 5$\sim$500-fold enhancement to the molecular diffusion.  Moreover, the computed diffusivity and particle displacement distributions are consistent with observations in several controlled experiments on biological fluids \citep{Leptos2009, Drescher2009,Pushkin2013}.  Admittedly, this simplified model does not account for some key characteristics of marine animals and environment, such as schooling, wake turbulence and vertical stratification.  Nonetheless, it serves as a good starting point to study the problem of biogenic mixing from a microscopic point of view and it accurately describes related phenomena in simpler settings.

This manuscript extends the model by Thiffeault and Childress by including simple interactions between swimmers.  Following the method of images for a potential flow past two cylinders first proposed by \cite{Carpenter1958} and later generalised to the case with more cylinders \citep{Dalton1971}, we use analytic streamfunctions for a hierarchy of doublets to compute the drift displacement induced by two or three cylinders moving synchronously.   In these potential flows, as well as low Reynolds number flows  generated by force-free swimmers, the inverse quadratic decay of far-field velocity guarantees that the squared displacement integrated over all possible impact parameters converges and thus we obtain the effective and enhanced scalar diffusivity.   We find that with just two cylinders, different configurations (in-school separation and inclination) produce nontrivial and nonlinear enhancement to previous results for non-interacting cylinders.  There are two distinct contributing mechanisms highlighted by opposite parameter dependences of the effective diffusivity.  We give a physical explanation to the chasing case (zero inclination) first and then examine the `active regions' in the parameter plane since they have close connections with the mixing enhancement for non-chasing formations.   While the methodology can be extended to study more cylinders with arbitrary positioning and asynchronous swimming with similar but much more tedious calculations, we are motivated by addressing the schooling effects in this simple model and therefore we restrict our discussion to the cases of two synchronously swimming cylinders.  

The manuscript is organised as follows:  Section \ref{sec:review} is a review of the model of random stirring by multiple bodies and the formula for effective diffusivity. In Section \ref{sec:images} we apply the method of image doublets and derive the formulas for the streamfunctions for a potential flow past two or three cylinders; In Section \ref{sec:results} we show the results for the effective diffusivity and investigate its dependence on configuration parameters.  Section \ref{sec:symmetry} takes a detailed look at the active mixing regions in a parameter plane and particle trajectories under schools with non-zero inclination.  We further make an exploratory attempt at the mixing effects of schools of three or more cylinders in Section \ref{sec:threecyl}.  Finally, we discuss the results and future directions in Section \ref{sec:discuss}.

\section{Stochastic hydrodynamic model\label{sec:review}}
Consider a passive particle submerged in an inviscid fluid in two dimensions.  A classical problem in hydrodynamics is the potential flow past a cylinder moving along a straight line and the explicit formula for the 2D streamfunction is available \citep{Maxwell1869}.  Consequently, the drift experienced by the particle can be readily computed by integrating the velocities in time.  It was shown by \cite{Thiffeault2010b} that the total particle displacement due to infrequent encounters with a dilute suspension of cylinders swimming in random directions can be modelled by the linear superposition
\begin{equation}
  \xv(t)= \xv_0 + \sum_{k=1}^{\Nenc(t)} \Delta_\pal(\ai_k,\bi_k) \, \ruv_k
  \label{eq:partdisp}
\end{equation}
where $\xv(t)$ is the particle displacement vector at time $t$, $\xv_0$ is its initial position, $(\ai_k,\bi_k)$ are the impact parameters imposed by the $k^{th}$ encountered swimmer who moves for a fixed distance $\lambda$ in the random direction $\ruv_k$ and $\Nenc(t)$ is the number of encounters as a function of time.  With proper averaging, the effective diffusivity of the scalar field is
\begin{equation}
\kappa:= \frac{\<\lvert\xv(\time)-\xv_0\rvert^2\> }{4t}=
  \frac{2\Uc\nd}{\pal}\int_0^\infty \int_{-\infty}^{\infty}
  \Delta_\pal^2(\ai,\bi)\db\da
  \label{eq:kappadef}
\end{equation}
where $\Uc$ is the  constant speed of the cylinders and $\nd$ is the number density of the swimmers.   To compute the individual drift $\Delta_\pal$, one only needs to differentiate the streamfunction of the potential flow past a cylinder and integrate the velocities in time.  When the swimming distance $\pal$ is much larger than the cylinder size $\lsc$ and can be assumed to be infinite, \cite{Thiffeault2010b} obtained
\begin{equation}
\kappa_{\text{s}}\approx1.19\,\Uc\nd\,\lsc^3.
\label{eq:kappas}
\end{equation}

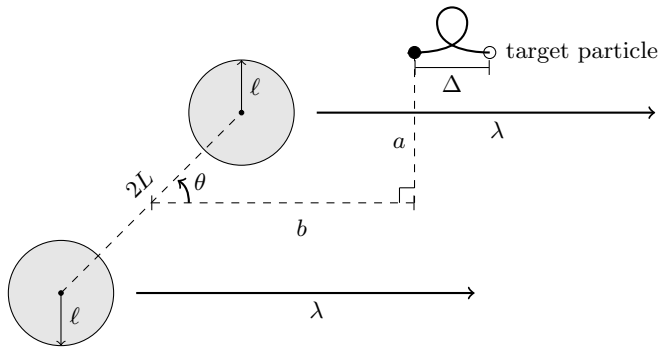
\begin{figure}
\centering
\vskip 1.5em
\begin{tikzpicture}
%
\draw [thick,<-] (1.35,.3) to [out=-30,in=90] (1.5,0);
\node [above] at (1.65,0.05) {$\theta$};
\draw [fill=black!10] (2.2,1.2) circle [radius=0.7];
\draw [fill] (2.2, 1.2) circle [radius=0.03];
\draw [->] (2.2,1.2) -- (2.2,1.9);
\node [right] at (2.2,1.55) {$\ell$};
\draw [->][thick] (3.2,1.2) -- (7.7,1.2);
\node [below] at (5.6,1.2) {$\lambda$};
\draw [fill=black!10] (-.2,-1.2) circle [radius=0.7];
\draw [fill] (-.2, -1.2) circle [radius=0.03];
\draw [->] (-.2,-1.2) -- (-.2,-1.9);
\node [right] at (-.2,-1.5) {$\ell$};

\draw [dashed] (-.2,-1.2) -- (2.2,1.2);

\node [above,rotate=45] at (1,.1) {$2L$};

\draw [->][thick] (.8,-1.2) -- (5.3,-1.2);
\node [below] at (3.2,-1.2) {$\lambda$};

\draw [fill] (4.5,2) circle [radius=0.08];
\draw (5.5,2) circle [radius=0.08];
\node [right] at (5.6,2) {target particle};
\draw [thick] (4.5,2) to [out=0,in=270] (5.2,2.4)  to [out=90,in=360] (5,2.6)
        to [out=180,in=90] (4.8,2.4) to [out=270,in=180] (5.5,2);
\draw [|-|] (4.5,1.8) -- (5.5,1.8);
\node [below] at (5,1.8) {$\Delta$};
\draw [dashed,|-] (4.5,2) -- (4.5,0);
\node [left] at (4.5,.8) {$a$};
\draw (4.3,.2) -- (4.5,.2); 
\draw (4.3,.2) -- (4.3,0);
\draw [dashed,|-|] (1,0) -- (4.5,0);
\node [below] at (3,-.1) {$b$};
\end{tikzpicture}
\caption{\label{fig:encdiag}The drift displacement $\Delta=\Delta_\pal(\ai,\bi,L,\theta)$ induced by a pair of schooling cylinders.  The impact parameters $\ai$ and $\bi$ denote the initial perpendicular distance and the horizontal distance, respectively, from the start of the particle trajectory (filled dot) to the midpoint between the cylinders.  The configuration parameter $\Lsep$ is defined as half of the separation between the cylinder centres whereas $\theta$ is the inclination angle from the swimming direction to the line connecting the centres.  The particle stops at the hollow dot when two cylinders finish their gliding of length $\pal$ with constant speed $\Uc$. }
\end{figure}

The accuracy of the approximation \eqref{eq:partdisp} relies on the dilute assumption: The number density $n$ has to be small so that the interaction between the cylinders in the potential flow is negligible.    In other words,  the drift imposed upon the passive particle at any instance of time comes predominantly from one swimmer.  This is violated when schools of multiple cylinders that are close to each other exist.  In this paper,  we look at the simplest nontrivial scenario of schooling:  The swimmers forms dilute, well-separated schools  while within each school, two identical cylinders stay close to each other and move synchronously with identical speed, duration and direction.   A diagram of each encounter between the passive particle and a school pair is illustrated in Fig. \ref{fig:encdiag}.  This is very similar to \cite{Thiffeault2010b} and \cite{Lin2011} with two more parameters introduced:  the separation between two cylinders, $2\Lsep$ ($\Lsep \ge \lsc$), and the inclination angle between the swimming direction and the line connecting the cylinder centres, $\theta$.

\section{Method of image doublets for potential flow past two cylinders \label{sec:images}}  

To extend the work of Thiffeault \& Childress we now derive the 
streamfunction for the potential flow generated by two cylinders moving synchronously as shown in Fig. \ref{fig:encdiag}.  It is easy to see that a simple superposition of two doublets would distort the impermeable boundaries from being circular, which inspired several methods to compensate for the inter-cylinder effects, including conformal mapping \citep{Crowdy2006}, elliptic function theory \citep{Johnson2004} and the method of image doublets
\citep{Carpenter1958, Dalton1971}. Here we demonstrate the method of image doublets due to its simplicity and derive the streamfunction and velocities from complex analysis.  

The basic idea is to construct an infinite series of image doublets with decreasing strength for each cylinder.  Within each series, the first, zeroth-order doublet represents the unperturbed cylinder and the $k^{th}$ ($k\ge 1$) image offsets the boundary distortion caused by $(k-1)^{th}-$order image in the other series.   Finally, the total complex potential is simply the sum of a uniform flow and two series
\begin{equation}
w=\phi+i\psi=-\Uc z+\sum_{k=0}^\infty w_{1,k}+\sum_{k=0}^\infty w_{2,k}.
\label{eq:potw}
\end{equation}
Here the uniform flow at infinity is moving from right to left and the convergence of these series is guaranteed by the decay of the doublet strength in each series. 

Next we derive the formulas for the doublets $w_{1,k}$ and $w_{2,k},\, k=0,1,\dots$.   This is equivalent to the determination of the position and the strength for each doublet.  Without loss of generality, for each encounter illustrated in Fig. \ref{fig:encdiag} we set up a co-moving, complex $z-$plane with the midpoint between the two cylinders being the origin, and with the swimming direction being the positive real ($x$) axis.   

It is known that for the zeroth-order doublets that model the potential flow past a cylinder \citep{Acheson1990}
\begin{equation}
w_{j,0}=-\frac{U\lsc^2}{z-z_{j,0}},\quad j=1,2
\label{eq:singcyl}
\end{equation}
with $z_{j,0}=(-1)^{j-1}\Lsep\, \ee^{i\theta}=\pm \Lsep(\cos\theta+i\sin\theta)$ as the coordinates of the cylinder centres that are symmetric with respect to the origin moving with the cylinders.  For the next order, since the image doublet $w_{1,1}$ offsets the circular boundary perturbation induced by doublet $w_{2,0}$ (within cylinder 2) around doublet $w_{1,0}$ (within cylinder 1), the following restriction should be imposed on the imaginary parts of the doublet potentials:
\begin{equation}
\Imp(w_{1,1}+w_{2,0})=-\Uc\lsc^2 \Imp\Big(\frac{s_{1,1}}{z-z_{1,1}}+\frac{1}{z-z_{2,0}}\Big)=\text{constant}
\label{eq:consimag}
\end{equation}
in which $s_{1,1}$ and $z_{1,1}$, the relative strength and position of the image doublet respectively, are chosen as follows \citep{Carpenter1958}:
\begin{eqnarray}
s_{1,1}&=&-\frac{\lsc^2\ee^{2i\theta}}{|z_{1,0}-z_{2,0}|^2}=-\frac{\lsc^2\ee^{2i\theta}}{4\Lsep^2},\\
 z_{1,1}&=&z_{1,0}+\frac{\lsc^2}{\;\overline{z_{2,0}-z_{1,0}}\;}=\ee^{i\theta}\Big(\Lsep-\frac{\lsc^2}{2\Lsep}\Big)
\label{eq:strengthandpos}
\end{eqnarray}
where $\overline{\;\cdot\;}$ denotes complex conjugacy.  Notice how the strength of the image decays according to an inverse square law for variable $\Lsep$ and how it lies on the line connecting two cylinder centres.  Similarly, the formula for $w_{2,1}$, the image doublet that restores the boundary distortion around cylinder 2 by $w_{1,0}$ can be derived. 

To summarise, the first order image doublets in the complex potential (\ref{eq:potw}) are
\begin{equation}
w_{j,1}=-\frac{\Uc\lsc^4e^{2i\theta}}{4\Lsep^2\Big[z+(-1)^j\ee^{i\theta}\Big(\Lsep-\dfrac{\lsc^2}{2\Lsep}\Big)\Big]},\qquad j=1,2.
\label{eq:w11}
\end{equation}
In fact, all higher order image doublets can be derived as above to balance corresponding lower order images in the same inductive fashion but with more tedious details.  However, as we will see in the next sections, first and second order images are sufficient for the purpose of computing effective diffusivity.  Furthermore, this procedure can be readily generalised for cylinders of different sizes and for more than two cylinders \citep{Dalton1971}.  We elect to postpone the discussion for these scenarios to future work since preliminary results show that more complicated configurations do not lead to significantly new phenomena in current context.

It should also be noted that the requirement (\ref{eq:consimag}) only enforces that the circular boundaries of the cylinders are impermeable streamlines.  The constant on the right hand side is generally nonzero and therefore the cylinders may be subject to lift and drag forces.  For more detailed derivations readers can refer to \cite{Dalton1971}.  To study the schooling effects, here we assume that there are `internal' forces constantly exerted on the cylinder pair to maintain their relatively stationary positions.

\section{\label{sec:results}Results for the effective diffusivity}

\begin{figure}
  \centering
  \includegraphics[width=.66\columnwidth]{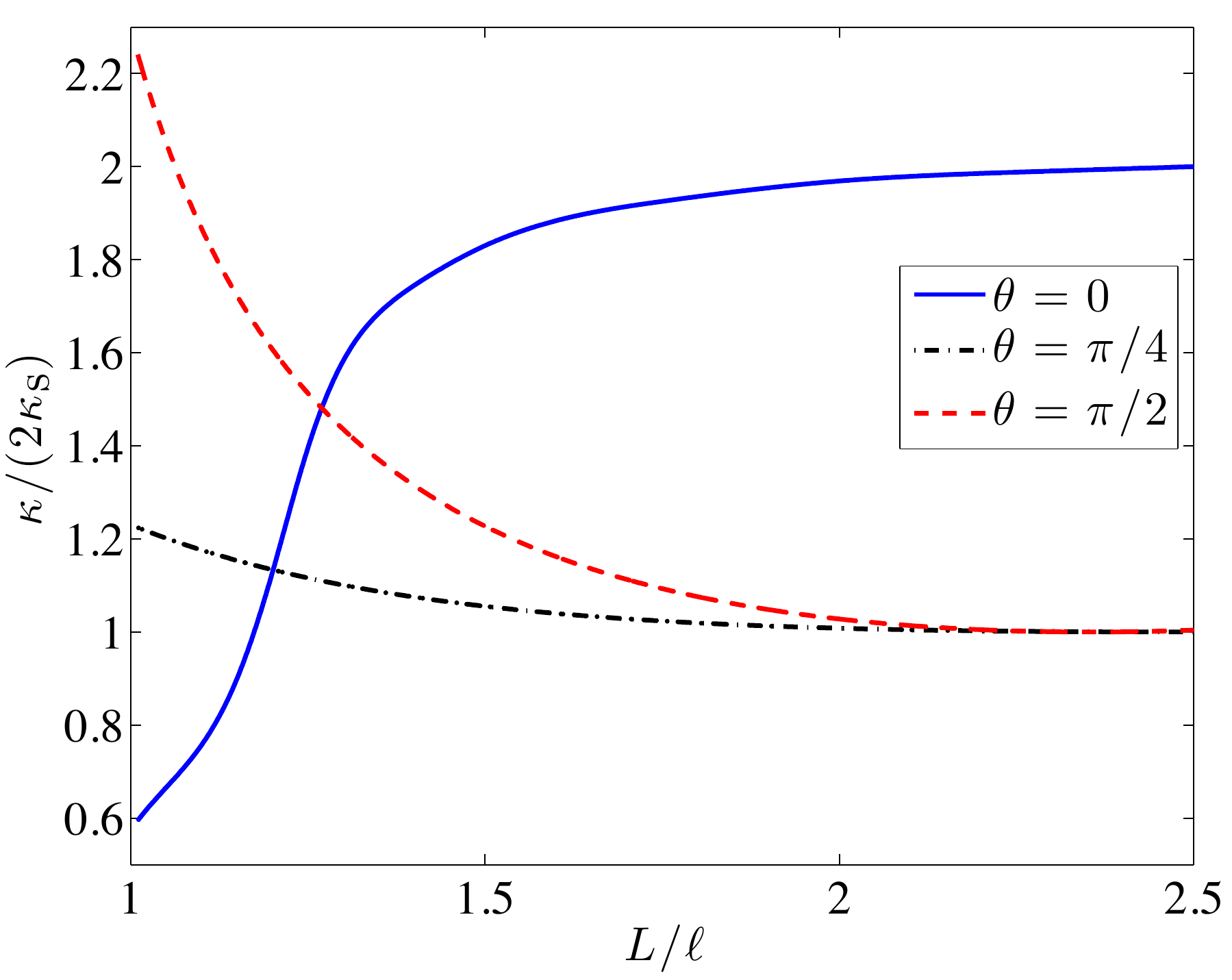}
  \caption{Normalised effective diffusivity $\kappa$ as a function of cylinder separation $\Lsep$. For different in-school formations, the dependence of the effective diffusivity on the separation parameter exhibits opposite behaviours: When $\chasing$, the diffusivity increases as the distance between two cylinder grows while for $\tilting$ and $\sweeping$ the monotonicity is reversed.  Here the number density $\nd=10^{-3}$, swimming speed $\Uc=1$ and the cylinder radius $\lsc=1$ are all kept constant.}
 \label{fig:kappa2}
\end{figure}

With the complex potential formulas presented above, we now compute the effective diffusivity $\kappa$ defined in (\ref{eq:kappadef}) for schools of two cylinders with two configuration parameters: cylinder separation $\Lsep$ and inclination $\theta$.   Fig. \ref{fig:kappa2} summarises the results for three typical formations: $\theta=0$ (`chasing'), $\pi/4$ (`tilting') and  $\pi/2$ (`sweeping').   The horizontal axis is the distance between cylinder centres normalised by the cylinder radius;  the vertical axis is the effective diffusivity normalised by twice the reference value (\ref{eq:kappas}).   The factor $2$ is introduced to highlight the nonlinearity in the schooling enhancement compared with simply doubling the swimmer density in the original model.  These curves are numerically generated by truncating each of the two series in the potential (\ref{eq:potw}) to three terms and with constants $\Uc=\lsc=1$, $\nd=10^{-3}$ and $\lambda=100$.   It has been verified for a wide range of  parameter settings that including more terms can change the result no more than $1\%$ due to the fast decay of image doublets.  In fact, keeping only two terms in each series, namely, considering only the zeroth-order doublets $w_{j,0}$ and their first-order corrections  $w_{j,1}$, $j=1,2$, recovers more than $96\%$ of the effective diffusivity.   Although the truncation does result in slight distortions in the cylindrical boundaries of the swimmers and thus some error in computing the particle drift very close to the boundaries,  their contribution to the integrated effective diffusivity is negligible.

Here we observe two opposite behaviours  of the effective diffusivity as a function of the separation parameter:  For the chasing case ($\chasing$), $\kappa$ is a strictly increasing function of $\Lsep$ while it is a strictly decreasing function in tilting ($\tilting$) and sweeping ($\sweeping$) formations.   Moreover, sweeping schools yield a much bigger boost than tilting ones when $\Lsep$ is small.  In all three cases, the dependence is nonlinear in that $\kappa$ varies rapidly for $\Lsep/\lsc<2$ and approaches an asymptotic, constant value as $\Lsep$ gets large.

A straightforward intuition can be applied to explain the two limiting cases for large $\Lsep$:  When a school of two swimmers chase through the fluid with enough separation,  each encounter with the particle can be well approximated by two sequential kicks by the leading then the trailing cylinder from exactly the same direction.   This \emph{long-range superposition} doubles combined particle drift which implies a fourfold increase in the squared displacement $\Delta^2$ and consequently $\kappa \approx 4\kappa_{\text{s}}$.  On the other hand, in the tilting and sweeping cases where $\theta$ is significantly different from 0,  the two cylinders no longer cooperate with each other when they are far apart and they act on the particle independently.    The resulting effective diffusivity is then simply a linear extrapolation of the value from independent swimmers and therefore $\kappa \approx 2\kappa_{\text{s}}$.  In other words, the swimmers achieve no advantage by schooling together under these configurations.  Of course, this asymptotic analysis is only valid when $\Lsep$ is still small compared to the average distance between two schools so the dilute assumption is still accurate.

\begin{figure}
 \centering
 \includegraphics[width=.66\columnwidth]{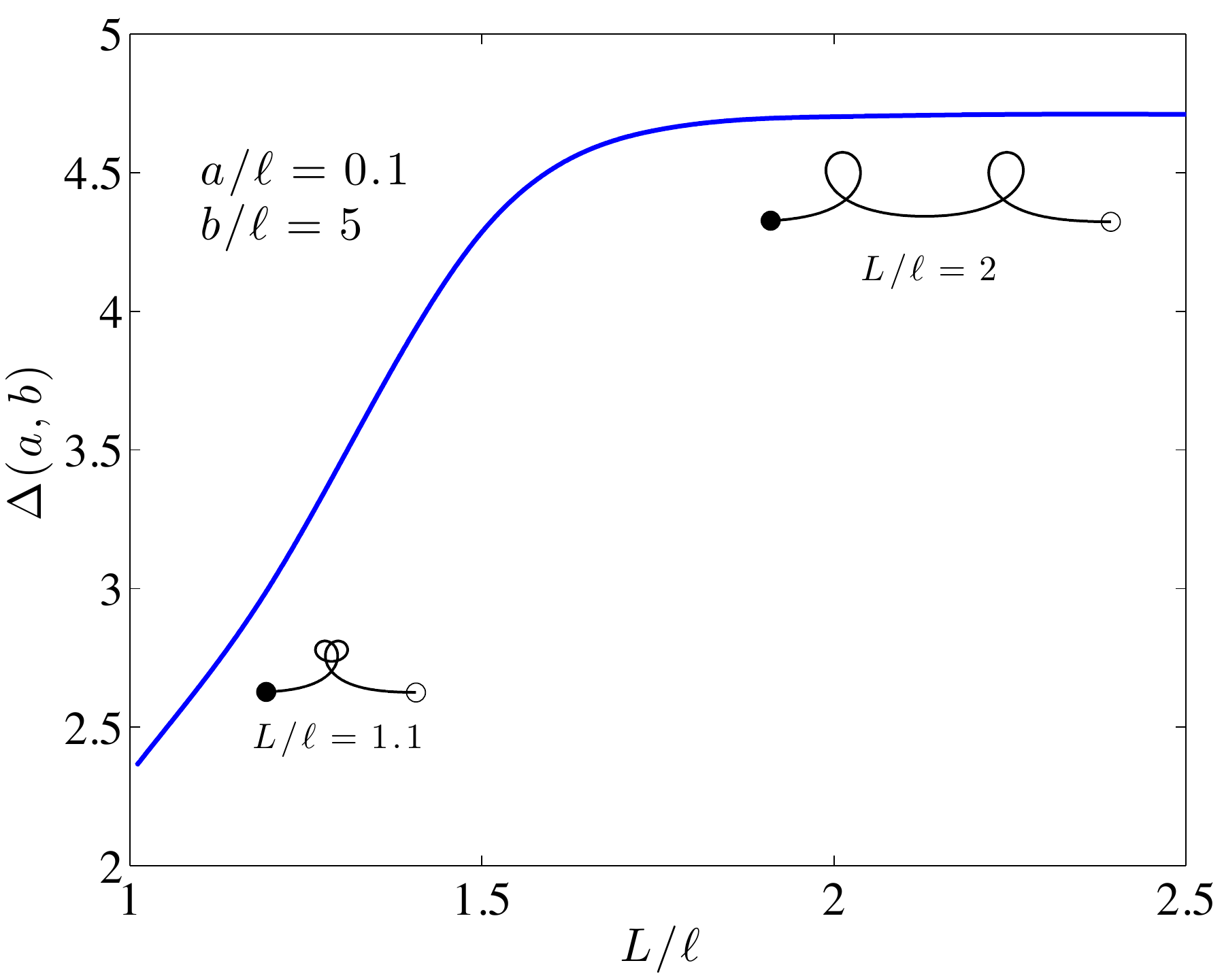}
  \caption{\label{fig:N=1_Delta}The effect of cylinder separation $\Lsep$ on the drift displacement $\Delta$ when $\chasing$.  Here the impact parameters $\ai/\lsc=0.1$ and $\bi/\lsc=5$ are fixed.  Two typical trajectories are overlaid. When $\Lsep$ is small, two cylinders operates nearly as one for the particle; When $\Lsep$ is large enough, two cylinders induce a drift displacement that almost doubles the single-cylinder value. }
\end{figure}

More careful investigation is required when $\lsc<\Lsep<2\lsc$.   For the chasing configuration, Fig. \ref{fig:N=1_Delta} illustrates how the particle drift depends on the separation $\Lsep$.  When the two cylinders are slightly apart ($\Lsep/\lsc =1.1$), the overlaid particle trajectory shows that the combined drift is only approximately equal to the single-cylinder value.   In this case, the schooling in fact \emph{suppresses} the mixing efficiency since doubling the number of swimmers by schooling does not double the effective diffusivity.  This regime can also be identified in Fig. \ref{fig:kappa2} in which part of the solid curve ($\chasing$) falls below unit value, namely, $\kappa/2\kappa_{\text{s}}<1$.  As $\Lsep$ grows,  the particle trajectory essentially becomes a superposition of two successive encounters as mentioned before and shown by the other overlaid trajectory in the figure ($\Lsep/\lsc=2$).   Therefore, schools in chasing formations can only enhance mixing significantly when the in-school separation is large enough.

\section{\label{sec:symmetry}Amplified active region and short-range coupling for $\theta\neq 0$}
\begin{figure}
  \centering
  \subfigure[]{ 
   	\includegraphics[width=.48\columnwidth]{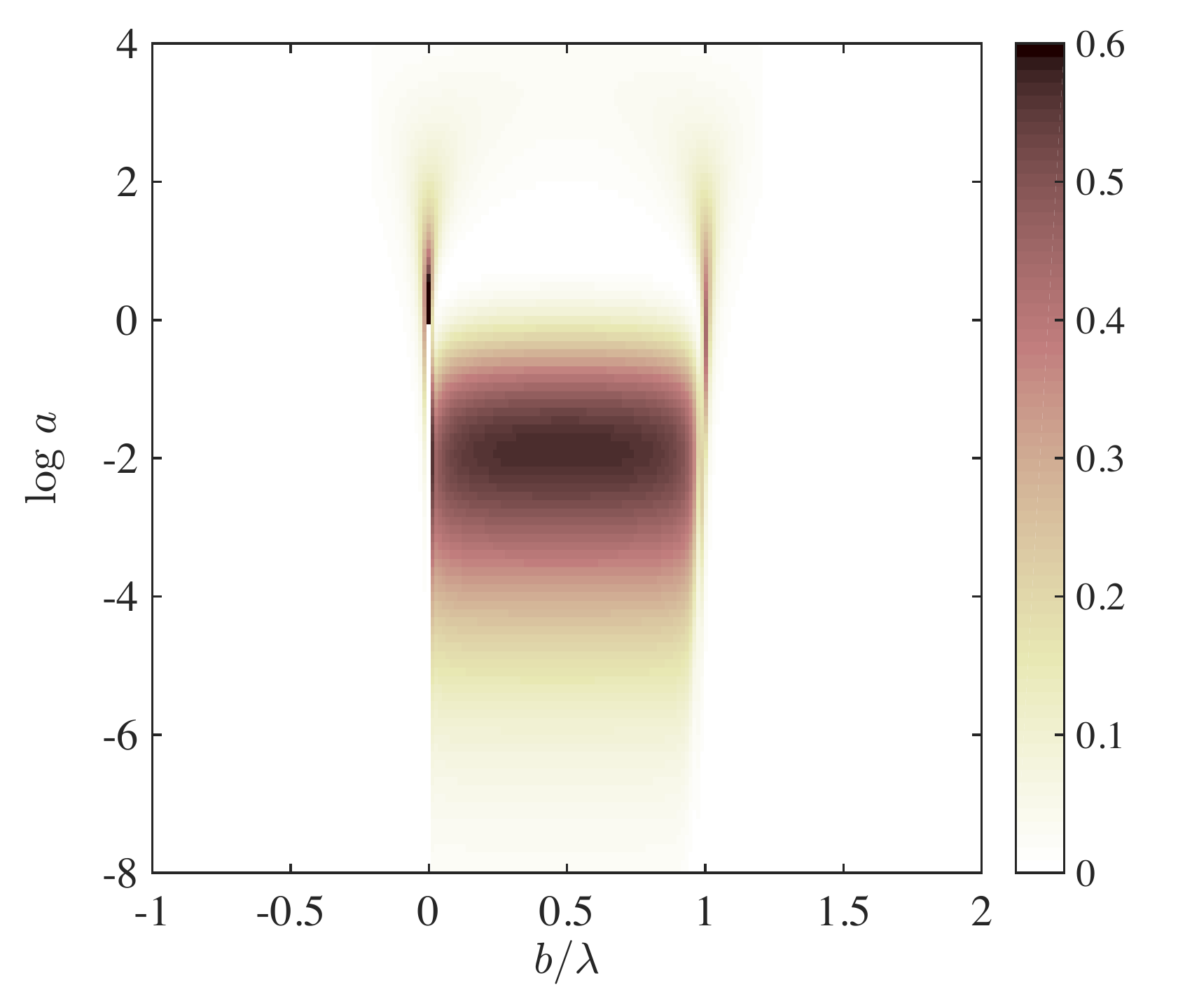}	
	\label{fig:dist_sing}
	}
   \subfigure[]{ 
   	\includegraphics[width=.48\columnwidth]{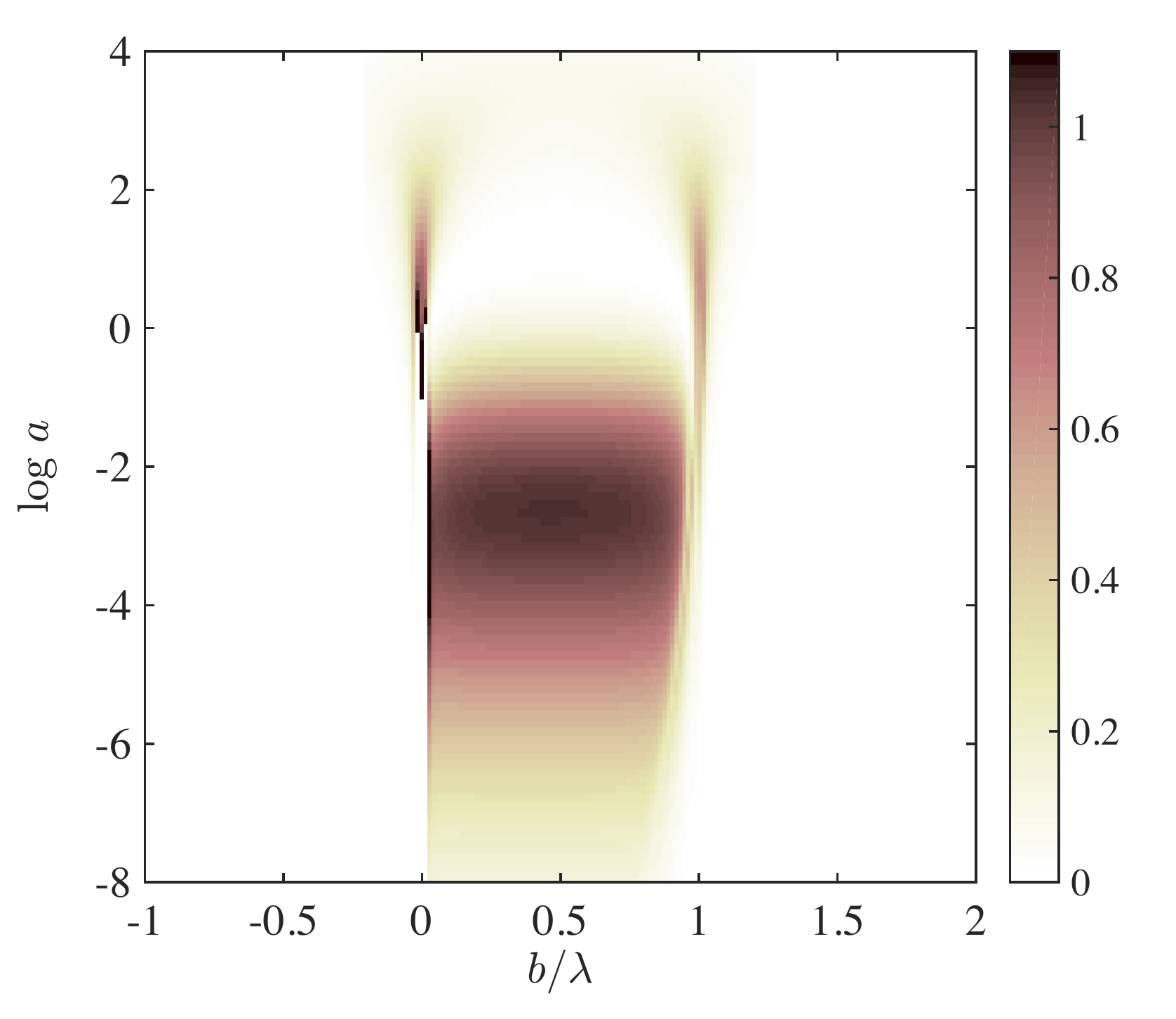}
	\label{fig:dist_chase}
	}
	\vskip -.3em
   \subfigure[]{ 
 	\includegraphics[width=.48\columnwidth]{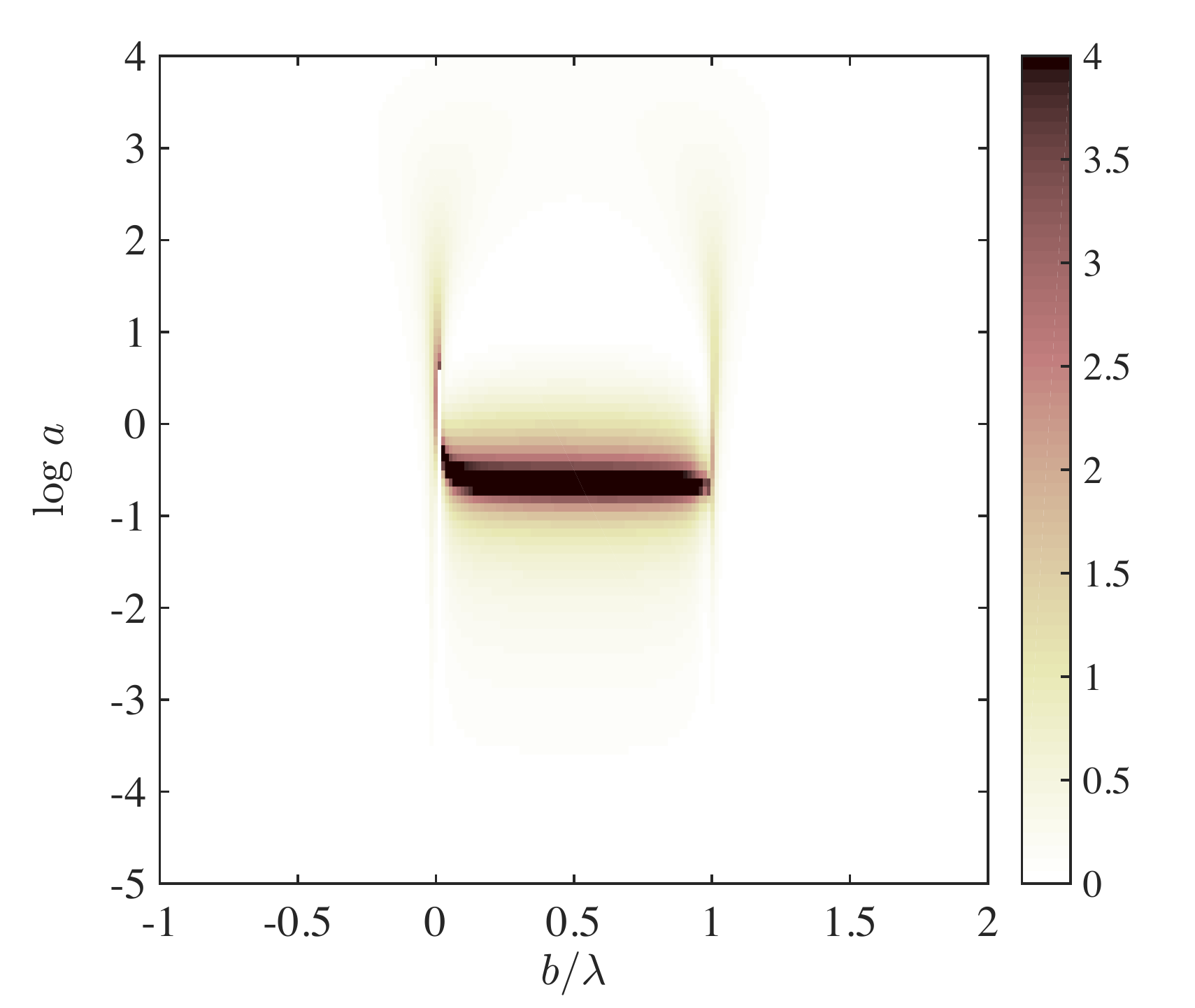}
	\label{fig:dist_tilt}
	} 
    \subfigure[]{
    	\includegraphics[width=.48\columnwidth]{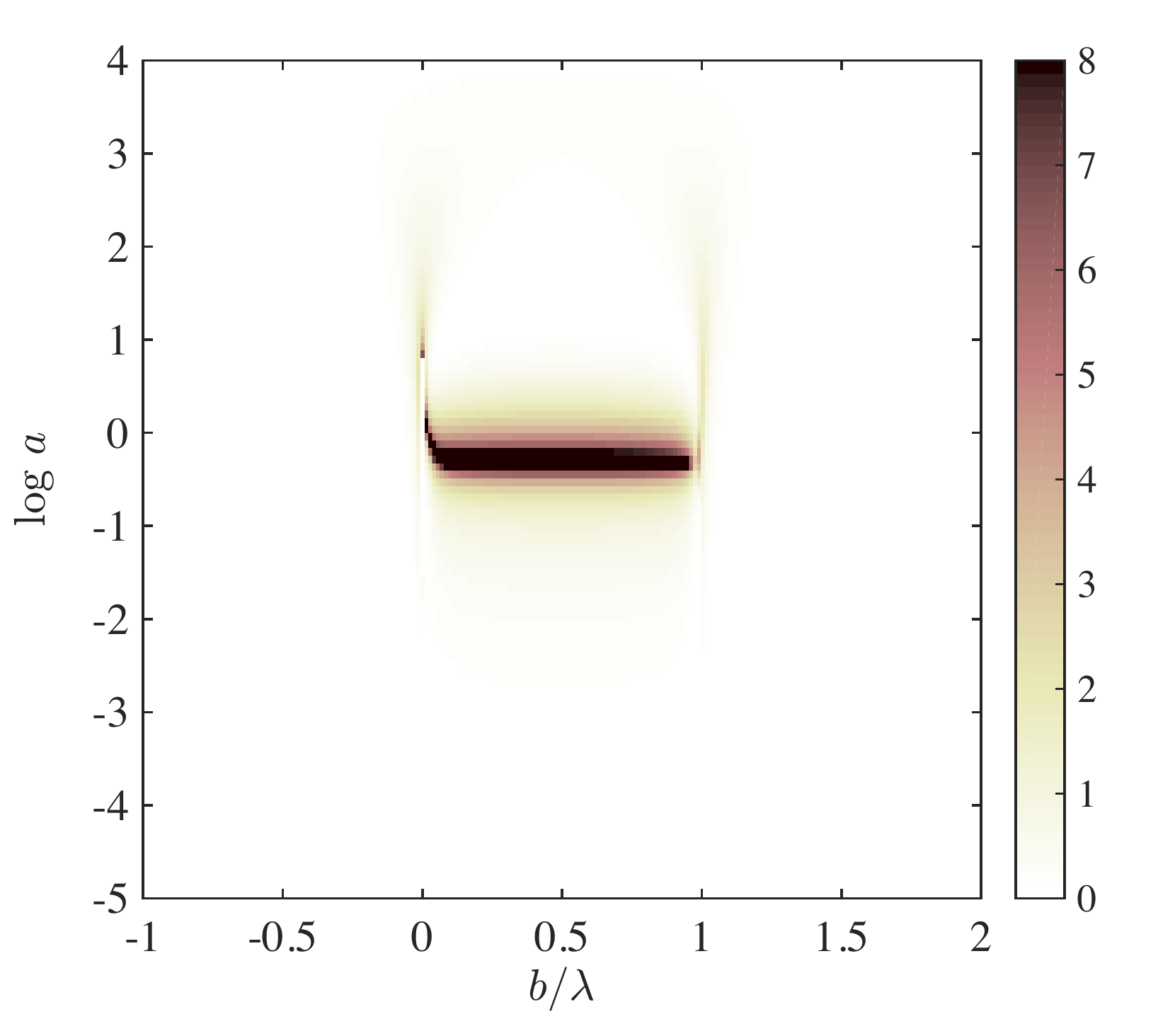}
	\label{fig:dist_sweep}
	}
  \caption{\label{fig:delta2}Distributions of integrand $\lsc^{-3}a\DisplacementSquared$ in (\ref{eq:transkappaint}) for (a) a single cylinder and for a two-cylinder school in three different configurations: (b) $\chasing$, (c) $\tilting$ and (d) $\sweeping$.  The significant values (plotted with darker shading) in cases (a) and (b) spread over $\ai\ll \lsc, \bi/\lambda\in(0,1)$ while they are greatly intensified and localised near $\ai\approx \lsc$ in cases (c) and (d). In all figures $\Uc=\lsc=1,\,\lambda=100$ and for all schooling configurations $\Lsep/\lsc=1.2$.}
  \end{figure}

Finally, we investigate the source of the nonlinearly enhanced diffusivity when $\theta\neq 0$.  In this section we fix $L=1.2$ without loss of generality.  First we examine the dominating contribution of the transformed $\kappa$ integral in the $\log(a/\lsc)$-$(b/\lambda)$ parameter plane
\begin{equation}
\lambda^{-1}\lsc^{-3}\int_{-\infty}^{\infty}\int_0^{\infty}\DisplacementSquared\,{\rm d}a\,{\rm d}b=\iint_{\mathbb{R}^2}\lsc^{-3}a\DisplacementSquared\,{\rm d}\log(a/\ell)\,{\rm d}(b/\lambda)
\label{eq:transkappaint}
\end{equation}
by comparing in Fig. \ref{fig:delta2} the distributions of the dimensionless integrand $\lsc^{-3}a\DisplacementSquared$  \citep{Lin2011} for the non-schooling case and for the three schooling formations.  The \emph{active region} in each panel is the area where the integrand value is significantly non-zero (shown in darker colour), or equivalently, is the set of $(a,b)$ such that $\lsc^{-3}a\DisplacementSquared\ge\gamma$ where the significance $\gamma$ is set to be $0.3$ here.  It should be noted that any reasonable choice of $\gamma$ supports the subsequent arguments.

We can see that the single-cylinder case and the chasing case (Fig. \ref{fig:dist_sing} and \ref{fig:dist_chase}) share a strong similarity: In both cases the integral is dominated by the active region $\log(\ai/\lsc)\lesssim -0.6, \,b/\lambda\in(0,1)$ which corresponds to the `head-on' collisions near $\ai=\Lsep \sin\theta=0$ (see Fig. \ref{fig:encdiag} with $\theta=0$) in physical coordinates.  And a chasing school produces longer drifts and thus a larger diffusivity.  By contrast, for tilting and sweeping schools (Fig. \ref{fig:dist_tilt} and \ref{fig:dist_sweep}) the dominating contributions come from a shifted and more localised region near $\log(\ai /\lsc)=0$, or $\ai =\Lsep\sin\theta=O(1)$.  As it moves away from $\ai\ll \lsc$,  the intensity in this region is greatly amplified to an extent that the integrated diffusivity achieves a nonlinear growth for $\theta\neq 0, \lsc<\Lsep<2\lsc$ as shown in Fig. \ref{fig:kappa2}.  In fact,  the integrand reaches its maximal value of 10 (or 16) when $\tilting$ (or $\sweeping$) for typical values of $\bi$ while the maximum is only 0.6 in the non-interacting case.   This magnification factor is much more than enough to compensate for the active area localisation when integrating $\kappa$ in (\ref{eq:transkappaint}) although, interestingly, this region actually \emph{expands} in the dimensional $a$-$b$ plane which will be mentioned again.  Both the magnification in strength and the expansion in dimensional space are much more significant for $\sweeping$ than for $\tilting$ and thus the advantage of sweeping over tilting entails.

The physical origin of the above analysis is attributed to the \emph{short-range coupling} via which tilting and sweeping schools with small $\Lsep$ produce large displacements and super-linearly enhanced diffusivity.  This coupling involves the simultaneously impacts on the particle from both cylinders when $\theta\neq 0$  which are in the same order of magnitude.   Furthermore, the coupling effect intensifies as $\theta$ increases towards $\pi/2$, or as $\Lsep$ decreases which agrees with what we have seen in Fig. \ref{fig:kappa2}.  However, this effect is generally negligible when $\theta=0$ since the subdominant, `farther' cylinder only provides a contribution in the order of $O(1/(1+2\Lsep/\lsc)^2)$ of what the dominant, `closer' one enforces.  

\begin{figure}
  \centering
  \includegraphics[width=.98\columnwidth]{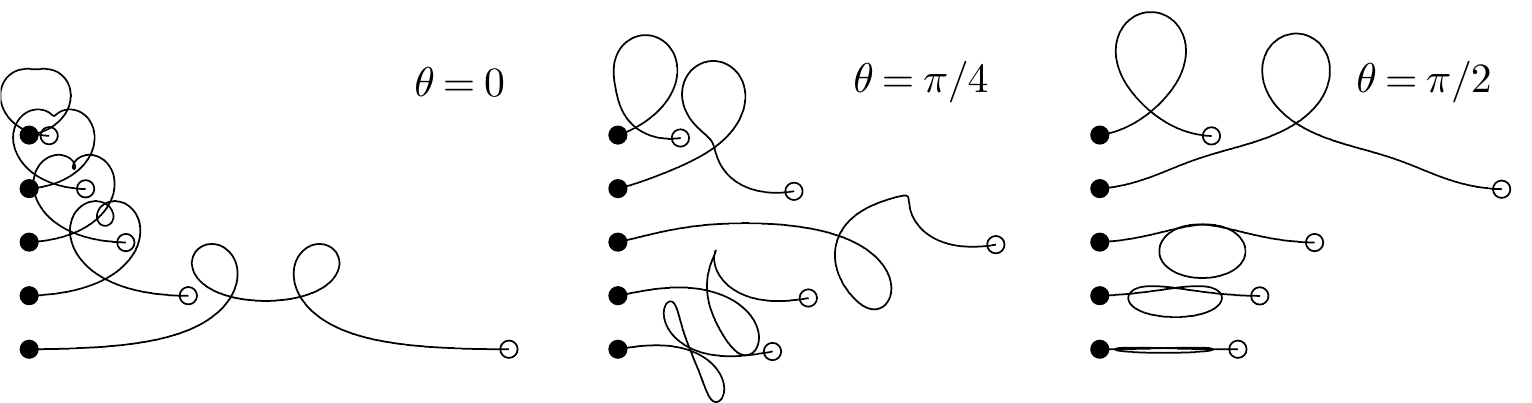}
  \caption{  \label{fig:trajs}Typical trajectories of a target particle in the fixed laboratory frame for different values of $\ai$ due to a two-cylinder school moving from left to right with $\Uc=\lsc=1$, $\lambda=100$ and $\bi/\lambda=0.3$. The initial positions of the particle are marked by solid dots and the final positions by hollow dots. In each panel, solid dots from top to bottom correspond to $\ai/\lsc = 1.2,\,0.9,\,0.6,\,0.3,\, 0.05$.  The cylinder separation is fixed in all three configurations at $\Lsep/\lsc=1.2$.  Note how the fore-aft symmetry of a trajectory is completely broken by a tilting school as seen in the centre panel.}
\end{figure}

Fig. \ref{fig:trajs} visualises this mechanism by superimposing typical trajectories for different values of $\theta$ and $\ai$.   The key observation one can make here is that the two cylinders within each of the $\theta\neq 0$ schools (centre and right panels) reinforce each other not just by invoking long drifts but also by slowing down the drift decay away from the head-on positions, $\ai=\Lsep\sin\theta$, such that particles with a broader range of impact parameters experience long drifts in comparison with the chasing case as noted in the previous paragraph.    It should also be noted that these enhanced drifts deviate from the close loops formed with same impact parameters under non-interacting settings~\citep{Lin2011, Pushkin2013b}. 

\section{\label{sec:threecyl}Schooling Effects of Three Cylinders}
Our next step is to explore the mixing effects of schools of three or more cylinders.  This is particularly natural in light of our motivating problem, ocean biogenic mixing, since in reality a fish school contains up to thousands of individuals. In theory, the method of image doublets can be applied to arbitrary number of cylinders in two dimensions. For example, for a school of three cylinders the construction of image doublet series analogous to Section \ref{sec:images} would start from three zeroth-order doublets, to 6 first-order image doublets (two first-order images within each zeroth-order doublet to balance the other two zeroth-order ones), to 12 second-order images and so on. 

\begin{figure}
\centering
\vskip 1.5em
\begin{tikzpicture}

\draw [fill=black!10] (-.2,0) circle [radius=0.6];
\draw [fill] (-.2,0) circle [radius=0.03];
\draw [fill=black!10] (-1.5,-.75) circle [radius=0.6];
\draw [fill] (-1.5, -.75) circle [radius=0.03];
\draw [fill=black!10] (-1.5,.75) circle [radius=0.6];
\draw [fill] (-1.5,.75) circle [radius=0.03];

\draw [dashed] (-.2,0) -- (-1.5,-.75);
\draw [dashed] (-.2,0) -- (-1.5,.75);
\draw [dashed] (-1.5,-.8) -- (-1.5,.8);
\node [below,rotate=30] at (-.75,-.43) {$2.4\ell$};
\node [left,rotate=90] at (-1.7,0.4) {$2.4\ell$};
\node [above,rotate=-30] at (-.75,.43) {$2.4\ell$};
\node [below] at (-.7,-1.8) {$\kappa/3\kappas=2.36$};

\draw [->][thick] (.8,0) -- (1.3,0);
\node [below] at (1.05,0) {$\lambda$};

\draw [fill=black!10] (3.2,0) circle [radius=0.6];
\draw [fill] (3.2,0) circle [radius=0.03];
\draw [fill=black!10] (4.5,-.75) circle [radius=0.6];
\draw [fill] (4.5, -.75) circle [radius=0.03];
\draw [fill=black!10] (4.5,.75) circle [radius=0.6];
\draw [fill] (4.5,.75) circle [radius=0.03];

\draw [->][thick] (5.3,0) -- (5.8,0);
\node [below] at (5.55,0) {$\lambda$};
\node [below] at (4.1,-1.8) {$\kappa/3\kappas=2.36$};

\draw [fill=black!10] (7.75,-.7) circle [radius=0.6];
\draw [fill] (7.75,-.7) circle [radius=0.03];
\draw [fill=black!10] (9.25,-.7) circle [radius=0.6];
\draw [fill] (9.25, -.7) circle [radius=0.03];
\draw [fill=black!10] (8.5,.6) circle [radius=0.6];
\draw [fill] (8.5,.6) circle [radius=0.03];

\draw [->][thick] (10,0) -- (10.5,0);
\node [below] at (10.25,0) {$\lambda$};
\node [below] at (8.8,-1.8) {$\kappa/3\kappas=2.52$};

\end{tikzpicture}
\caption{\label{fig:triangles}Three configurations for a three-cylinder school. In each configuration the cylinders form an equilateral triangle with $\Lsep$ fixed at $1.2\ell$ and three cases differ in the triangle orientation relative to the moving direction.}
\end{figure}
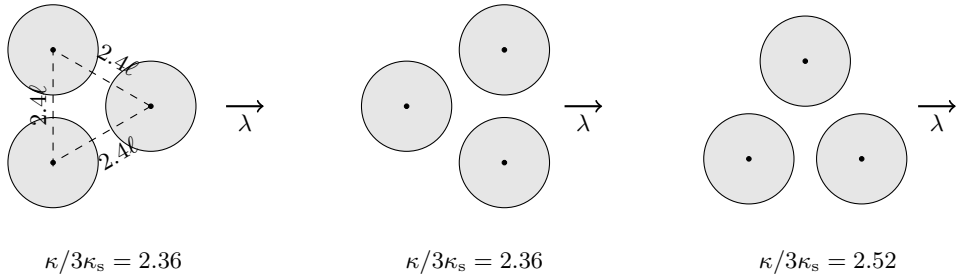

We have therefore conducted a preliminary study on triple-cylinder schools as extensions to two-cylinder formations.  It is to be expected that the effective diffusivity would have more subtle dependence on the parameters due to the extra degrees of freedom characterising in-school configuration.   Again, we truncated the infinite series to first-order images and more than 95 \% of the mixing effects was captured in all of our test cases. In particular, in Figure \ref{fig:triangles} we show three representative configurations for an equilateral, three-cylinder school with side lengths $2\Lsep=2.4\ell$ and these configurations all produce super-linear mixing enhancement ($\kappa/3\kappas>1$).   In fact, the large-$\Lsep$ limit of the normalised effective diffusivity, $\kappa/3\kappas$, can go up to 3 in contrast to 2 in the two-cylinder cases shown in Fig. \ref{fig:kappa2}.  Not surprisingly, this can be achieved by a chasing formation in which three cylinders align and move along a straight line just as the two-cylinder case.  Figure \ref{fig:kappa3} summarises the effective diffusivity as a function of the equal separation $2\Lsep$ under two types of formation and we see very similar curves as in Fig. \ref{fig:kappa2}.   

\begin{figure}
\centering
\vskip 1.5em
\begin{tikzpicture}

\draw [fill=black!10] (1,-1) circle [radius=0.6];
\draw [fill] (1,-1) circle [radius=0.03];
\draw [fill=black!10] (1,1) circle [radius=0.6];
\draw [fill] (1,1) circle [radius=0.03];
\draw [fill=black!10] (-1,-1) circle [radius=0.6];
\draw [fill] (-1, -1) circle [radius=0.03];
\draw [fill=black!10] (-1,1) circle [radius=0.6];
\draw [fill] (-1,1) circle [radius=0.03];

\draw [dashed] (1,-1) -- (1,1);
\draw [dashed] (1,1) -- (-1,1);
\draw [dashed] (-1,1) -- (-1,-1);
\draw [dashed] (-1,-1) -- (1,-1);
\node [above] at (0,1) {$2\Lsep$};
\node [below] at (0,-1) {$2\Lsep$};
\node [left,rotate=90] at (-1.25,0.3) {$2\Lsep$};
\node [right,rotate=90] at (1.25,-.28) {$2\Lsep$};

\draw [->][thick] (2,0) -- (3,0);
\node [below] at (2.5,0) {$\lambda$};

\end{tikzpicture}
\caption{\label{fig:4cylinders}Square formation for a four-cylinder school. The normalised diffusivity is found to approach 2.45 as $\Lsep$ grows. }
\end{figure}
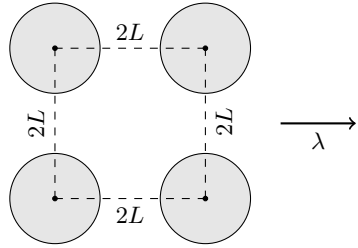

Regardless of whether it is plausible to extrapolate this into a quadratic rule, namely,
\begin{equation}
\max\frac{\kappa}{\kappas}=N^2
\label{eq:squarebound}
\end{equation}
over all possible in-school configurations where $N$ is the school size, or whether the chasing formation is always the optimal,  a more detailed study would be required.   As suggestive as Fig. \ref{fig:kappa2} and \ref{fig:kappa3} may seem, we are yet to conclude that the maximal diffusivity is indeed 3 and that it is uniquely realised by a chasing formation attributing to the difficulty in exhausting and parametrising all possible three-cylinder configurations, such as triangles with unequal side lengths, not to mention the resulting formulas for image doublets that are much more tedious and less tractable.  Complications of this nature are further multiplied when we consider four or more cylinders within one school.  However, the above results and analyses do provide some insights, at least under the settings of a planar, potential flow, to how in-school interactions between cylinders can translate to enhanced mixing.    In both two-cylinder and three-cylinder settings, the long-range superposition (successive kicks) and short-range coupling (amplified active region) have been shown to be effective mechanisms in terms of producing super-linear enhancements.    With the inclusion of even more cylinders, we conjecture that these two would remain the dominant contributors although their simultaneous operation would definitely add to the nonlinearity and thus the complexity of the problem.  For example, we compute the normalised diffusivity for the four-cylinder configuration illustrated in Figure \ref{fig:4cylinders} and found that $\kappa/4\kappa_s\to 2.45$ as $L\to\infty$ which significantly deviates from unity seen for non-chasing two- or three-cylinder formations.  Due to the scope of this manuscript, we elect to postpone further study along this direction to future work.

\begin{figure}
  \centering
 \includegraphics[width=.68\columnwidth]{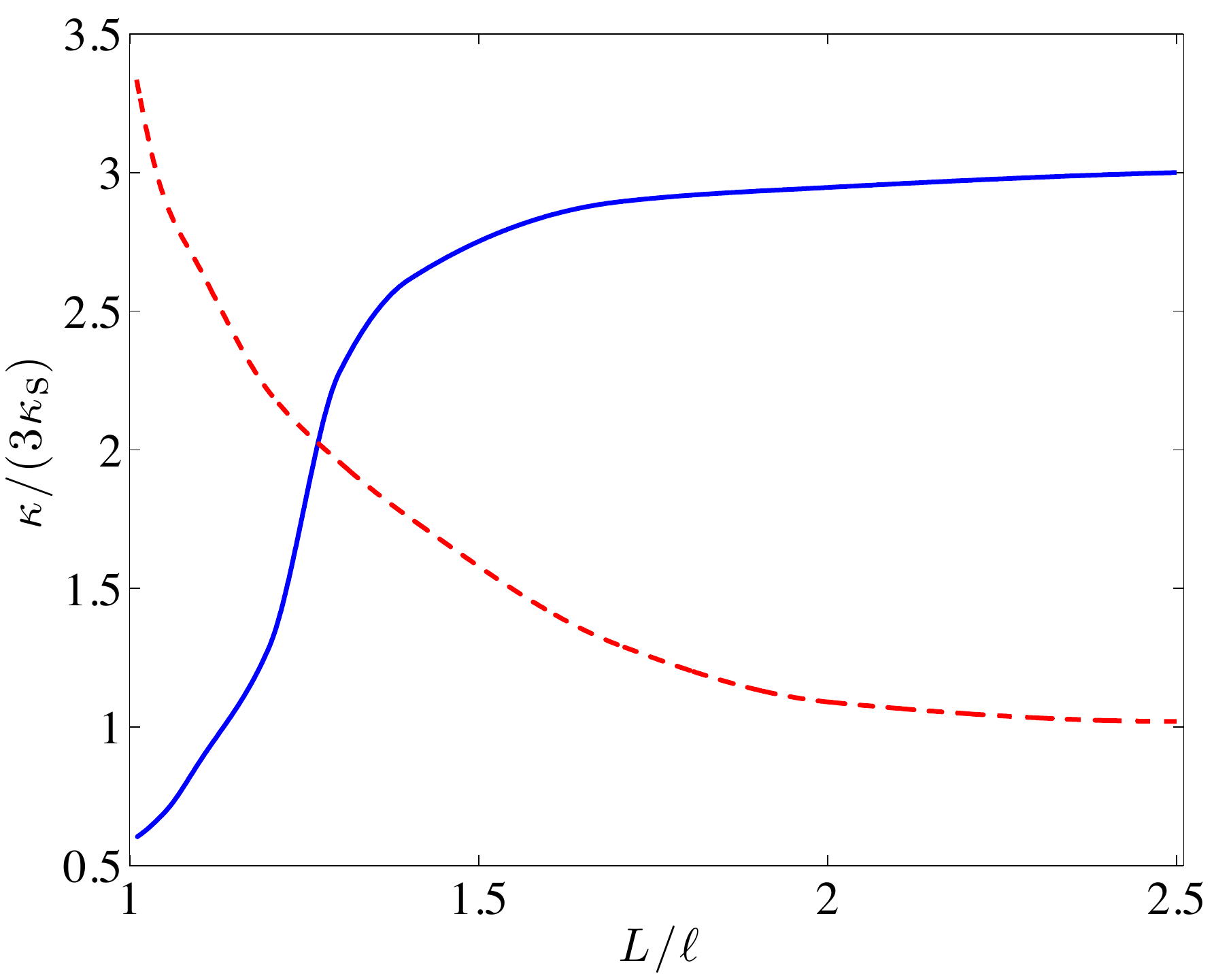}
 
 \vskip -21.2em
 \hskip 16em
 \begin{tikzpicture}
\draw [fill=black!10] (-.2,0) circle [radius=0.2];
\draw [fill] (-.2,0) circle [radius=0.03];
\draw [fill=black!10] (.6,0) circle [radius=0.2];
\draw [fill] (.6,0) circle [radius=0.03];
\draw [fill=black!10] (1.4,0) circle [radius=0.2];
\draw [fill] (1.4,0) circle [radius=0.03];

\draw [dashed] (-.2,0) -- (.6,0);
\draw [dashed] (.6,0) -- (1.4,0);
\node [above] at (.2,0) {$2\Lsep$};
\node [above] at (1,0) {$2\Lsep$};
\draw [->][thick] (2,0) -- (2.5,0);
\node [above] at (2.25,0) {$\lambda$};

\draw [fill=black!10] (.7068,-3.6) circle [radius=0.2];
\draw [fill] (.7068,-3.6) circle [radius=0.03];
\draw [fill=black!10] (1.4,-3.2) circle [radius=0.2];
\draw [fill] (1.4,-3.2) circle [radius=0.03];
\draw [fill=black!10] (1.4,-4) circle [radius=0.2];
\draw [fill] (1.4,-4) circle [radius=0.03];

\draw [dashed] (.7068,-3.6) -- (1.4,-3.2);
\draw [dashed] (.7068,-3.6) -- (1.4,-4);
\draw [dashed] (1.4,-3.2) -- (1.4,-4);
\node [below,rotate=-30] at (1.05,-3.85) {$2\Lsep$};
\node [left,rotate=90] at (1.7,-3.25) {$2\Lsep$};
\node [above,rotate=30] at (1.05,-3.3) {$2\Lsep$};

\draw [->][thick] (2,-3.6) -- (2.5,-3.6);
\node [above] at (2.25,-3.6) {$\lambda$};

\end{tikzpicture}
 
 \vskip 7em
 
  \caption{\label{fig:kappa3}Normalised effective diffusivity as a function of cylinder separation $\Lsep$ for three-cylinder schools in two types of formation in which the behavior of the effective diffusivity follows similar patterns as in Figure \ref{fig:kappa2}.}
  \end{figure}
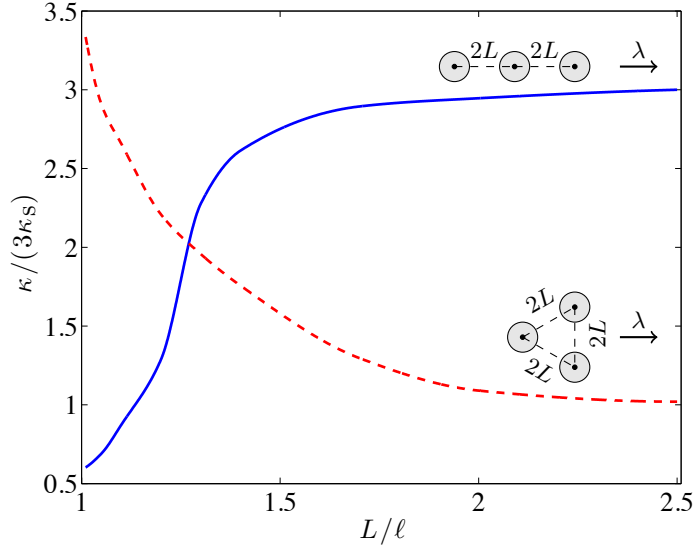

\section{\label{sec:discuss}Discussions and Future Work}

In this work, we focus on the mixing efficiency of a schooling pair of two cylinders moving synchronously in a potential flow and conclude that chasing and sweeping schools are generally more advantageous than tilting ones.  Furthermore, we illustrate the two physical mechanisms, long-range superposition ($\chasing$) and short-range coupling ($\theta\neq 0$), through which these schools generate large drift displacements and and thus a super-linear growth in the effective mixing diffusivity compared with non-interacting cylinders.  Among others, we will improve our understanding to the impact of schooling in the context of ocean mixing and of other realistic scenarios in several aspects.   

First is to study schooling effects in other fluid regimes and geometries, such as Stokes flow, a classical model for slow moving bodies in viscous fluids, and other three-dimensional flows.   We limited our discussion to  two-dimensional schools due to the simplicity in the series representation of image doublets and in the physical explanation to the enhanced diffusion.  Under axisymmetric scenarios, \cite{Lin2011} considered the mixing effects of a dilute suspension of independent Stokesian squirmers~\citep{Blake1971, Lighthill1952} in three dimensions and established connections to various biofluids~\citep{Drescher2009, Guasto2010, Ishikawa2007b, Leptos2009}, along with analogous results for independent spheres in a 3D potential flow.  More complicated boundary conditions and the loss of axisymmetry generally pose substantial challenges to extend the current theoretical framework.  One way to explore schooling effects under those settings is to first study special, axisymmetric formations of three-dimensional schools for which one can still characterise the flow field with streamfunctions and singularities, although the construction of image doublets would be much more involved.  As a qualitative conjecture, we expect that the mixing enhancement would be weaker in 3D analogous to the comparison between cases in two and three dimensions documented in \cite{Thiffeault2010b} due to the extra dimension and faster velocity decay.  On the other hand, with viscosity introduced as a realistic fluid condition, we expect even higher mixing enhancement as a result of stronger coupling between swimmers. 

Alternatively, one should also include more swimmers within each school in the context of studying ocean biogenic mixing, since in reality a fish school contains up to thousands of individuals.  We have seen in Section \ref{sec:threecyl} how straightforward generalisations of the method of image doublets for three-cylinder schools yielded complicated formulas and incomplete results.  For even larger schools, we may need to resort to a simple, macroscopic effective model to somehow parametrise the microscopic in-school interactions.   Specifically, a school of small fishes can behave effectively like a large fish at ocean mixing which, unlike a small fish, exhibits considerable mixing efficiency~\citep{Visser2007}.   What we can reasonably conjecture for a synchronised school of thousands of individual swimmers is that under certain formations (relatively small swimmer separations {\em perpendicular} to the swimming direction as in sweeping two-cylinder schools, and not too small separation in the swimming direction as in chasing two-cylinder schools), two mixing-enhancing mechanisms discussed before would coexist and cooperate in a way that demands further investigation.   In fact, Fig. \ref{fig:kappa3} has shown that three-cylinder schools that possess dominant chasing or sweeping components do indeed produce greater enhancements.

Additionally, it would be interesting to be able to identify the critical inclination $\theta^*>0$, if it exists, at which the curves shown in Fig. \ref{fig:kappa2} bifurcate from the increasing behaviour for $\theta=0$ to the decreasing behaviour for $\theta=\pi/4$ as the separation $\Lsep$ grows.

\begin{acknowledgments}
The authors are grateful to Jean-Luc Thiffeault for illuminating discussions. ZL was supported by National Natural Science Foundation of China under grants 11201419, J1210038 and 11426235.
\end{acknowledgments}

\end{document}